\documentclass[sigconf, natbib=false]{acmart}

\usepackage{comment}
\usepackage{booktabs}
\usepackage{multirow}
\usepackage{siunitx}
\usepackage{subcaption}
\usepackage{graphicx}
\usepackage{listings}
\usepackage{xcolor}
\usepackage{fancyhdr}

\colorlet{punct}{red!60!black}
\definecolor{background}{HTML}{EEEEEE}
\definecolor{delim}{RGB}{20,105,176}
\colorlet{numb}{magenta!60!black}

\lstdefinelanguage{json}{
    basicstyle=\small\ttfamily,
    numbers=left,
    numberstyle=\scriptsize,
    stepnumber=1,
    numbersep=8pt,
    showstringspaces=false,
    breaklines=true,
    frame=lines,
    backgroundcolor=\color{background},
    literate=
     *{0}{{{\color{numb}0}}}{1}
      {1}{{{\color{numb}1}}}{1}
      {2}{{{\color{numb}2}}}{1}
      {3}{{{\color{numb}3}}}{1}
      {4}{{{\color{numb}4}}}{1}
      {5}{{{\color{numb}5}}}{1}
      {6}{{{\color{numb}6}}}{1}
      {7}{{{\color{numb}7}}}{1}
      {8}{{{\color{numb}8}}}{1}
      {9}{{{\color{numb}9}}}{1}
      {:}{{{\color{punct}{:}}}}{1}
      {,}{{{\color{punct}{,}}}}{1}
      {\{}{{{\color{delim}{\{}}}}{1}
      {\}}{{{\color{delim}{\}}}}}{1}
      {[}{{{\color{delim}{[}}}}{1}
      {]}{{{\color{delim}{]}}}}{1},
}

\AtBeginDocument{%
  }

\copyrightyear{2022} 
\acmYear{2022} 
\setcopyright{acmlicensed}\acmConference[SCORED '22]{Proceedings of the 2022 ACM Workshop on Software Supply Chain Offensive Research and Ecosystem Defenses}{November 11, 2022}{Los Angeles, CA, USA}
\acmBooktitle{Proceedings of the 2022 ACM Workshop on Software Supply Chain Offensive Research and Ecosystem Defenses (SCORED '22), November 11, 2022, Los Angeles, CA, USA}
\acmPrice{15.00}
\acmDOI{10.1145/3560835.3564554}
\acmISBN{978-1-4503-9885-5/22/11}

\RequirePackage[
  datamodel=acmdatamodel,
  style=acmnumeric,
  ]{biblatex}

\newcommand{\toolname}[0]{GHAST}

\begin{document}

\title{Automatic Security Assessment of GitHub Actions Workflows}

\author{Giacomo Benedetti}
\email{giacomo.benedetti@dibris.unige.it}
\orcid{0000-0003-2609-6787}
\affiliation{%
  \institution{DIBRIS - University of Genoa}
  \city{Genova}
  \country{Italy}
}
\author{Luca Verderame}
\email{luca.verderame@dibris.unige.it}
\orcid{0000-0001-7155-7429}
\affiliation{%
  \institution{DIBRIS - University of Genoa}
  \city{Genova}
  \country{Italy}
}
\author{Alessio Merlo}
\email{alessio@dibris.unige.it}
\orcid{0000-0002-2272-2376}
\affiliation{%
  \institution{DIBRIS - University of Genoa}
  \city{Genova}
  \country{Italy}
}

\renewcommand{\shortauthors}{Giacomo Benedetti, Luca Verderame, \& Alessio Merlo}

\begin{abstract}
The demand for quick and reliable DevOps operations pushed distributors of repository platforms to implement \textsl{workflows}.
Workflows allow automating code management operations directly on the repository hosting the software.

However, this feature also introduces security issues that directly affect the repository, its content, and all the software supply chains in which the hosted code is involved in. 
Hence, an attack exploiting vulnerable workflows can affect disruptively large software ecosystems. 

To empirically assess the importance of this problem, in this paper, we focus on the de-facto main distributor (i.e., GitHub). We developed a security assessment methodology for GitHub Actions workflows, which are widely adopted in software supply chains.
We implemented the methodology in a tool (\toolname) and applied it on 50 open-source projects. 

The experimental results are worrisome as they allowed identifying a total of 24,905 security issues (all reported to the corresponding stakeholders), thereby indicating that the problem is open and demands further research and investigation.
\end{abstract}



\maketitle

\section{Introduction}
\label{sec:introduction}

Code repositories (also known as CRs) are a crucial part of every software development process.
They are used to store, share, distribute, and version software and its dependencies.
Indeed, code repositories are widely used in Software Supply Chains (SSCs) as they enable the storage of the core software as well as its distribution.
In detail, the software's coding and building phase integrate several pieces of code (e.g., modules and third-party libraries) hosted in separate CRs. To this aim, the corresponding SSC will include a set of CRs contributing to the final software.

Also, in the last years, the push for automation procedures of Continuous Integration/Continuous Delivery (CI/CD) led the distributors of the most common platforms (e.g., GitHub, GitLab, Bitbucket) to introduce engines for the execution of \emph{workflows}. A workflow is a sequence of actions aiming to automatize software's building, testing, and verification.
In practice, they allow automatizing CI/CD processes directly into the repository without relying on external services.
Workflows can be configured to run when manually triggered, at a scheduled time, or when a particular event on the repository occurs.

GitHub publicly released GitHub Actions (GHA) in 2019, and thanks to the provider's popularity, this service started to be widely used by developers.
In a nutshell, GHA consists of an API and a dedicated engine that allow users to define and execute workflows on their repositories. GHA engines support the execution of workflows on GitHub dedicated machines or self-hosted ones.

The ability of workflows to manage and modify the content of CRs makes them an appealing target for attackers. 
For instance, several technical reports~\cite{trendmicro, gitguardian} provided examples of attacks targeting GHA workflows to obtain control of GHA engines or inject malicious code into a repository.
Through the SSC, the compromised CR is able to affect the other nodes that rely on it, e.g., the final software importing the compromised code in its codebase.

To cope with the security implications of workflows and GHA, in this paper, we provide the following contributions.
First, we analyzed the GitHub security guidelines for hardening workflows, and we extracted a set of security constraints regarding confidential information, third-party workflows, permissions, and context variables. 
Then, we proposed a methodology to evaluate the security posture of GHA workflows, and we presented a prototype evaluation, called GitHub Action Security Tester (GHAST), based on the \textit{Sunset} SSC security framework \cite{Benedetti2022AliceEvaluation}. Finally, we conducted an in-the-wild experimental campaign on 50 GitHub - publicly available - repositories. The results allowed us to provide an overview of the security status of repositories.
From this evaluation, we analyzed the workflows of 646 unique repositories involved in the SSCs of all the 50 projects and identified 20 previously unknown security vulnerabilities and 24,885 security misconfigurations.

In the rest of the paper, we first provide the necessary background (Section~\ref{sec:background}) to understand the proposed contribution.
Then, we present our security assessment methodology to evaluate the security posture of workflows (Section~\ref{sec:contribution}). 
Moreover, we provide an implementation of the methodology (\toolname), and the techniques applied to rebuild the SSC and extract the required data in Section~\ref{sec:implementation}. 
Section~\ref{sec:results} is dedicated to the analysis of results obtained applying \toolname~in the wild against 50 open source projects.
Section~\ref{sec:related_work} discuss some related work.
Finally, Section~\ref{sec:conclusion} draws some conclusions and discusses some future extensions of this work.

\section{GitHub Actions: Background \& Security Issues}
\label{sec:background}
This section provides a brief description of the concepts discussed in our work.
In particular, we focus on GitHub Actions (GHA) and their core elements.
Then we discuss vulnerabilities affecting workflows, specifically GHA workflows, and how an attacker can exploit them to compromise code repositories in the SSC.

\subsection{GitHub Actions}
\label{subsec:github-actions}
GHA was introduced in 2019 with the aim to automate, customize, and execute software development workflows in code repositories.
GHA  attracted developers to ease automation routines in their projects independently of their size. At the time of writing (July 2022), the first 100 most starred repositories on GitHub adopted GHA.

GHA is an event-driven API provided by the GitHub platform to automate development workflows \cite{gh_actions}.
Among the different concepts introduced by GHA and required for their use, we give a brief explanation of the essential ones.
The GHA API enables the definition of workflows in one or more files in YAML format that need to be stored in the \texttt{.github/workflows/} directory of the target GitHub repository.

A workflow is composed of one or more jobs.
A job allows developers to define the environment and configuration where a sequence of tasks (namely, \emph{steps}) will run.
Each workflow can be associated with a list of events that trigger its execution. Examples of events include \texttt{pull request}, \texttt{push}, and \texttt{merge}. 
Runners are computing elements that host the execution of workflows for repositories. A Runner can be hosted both on GitHub dedicated servers and self-hosted machines.

At the start of each workflow run, GitHub automatically creates a unique \texttt{GITHUB\_TOKEN} to authenticate the request, granting each runner privileges to interact with the repository on behalf of GHA.
Administrators of the repository can set the permissions granted to the token to restrict access to specific resources or jobs. The default permissions can be either permissive or restricted. In the first case, the \texttt{GITHUB\_TOKEN} has full access to the resources of the repository, while - in the second case - the capabilities are limited to read the content of the repository \cite{token_auth}.

GHA offers developers artifacts to manage different aspects of workflow execution. 
Contexts are a way to access information about workflow runs, runner environments, jobs, and steps. Each context is an object that contains properties, which can be strings or other objects. Among the list of possible contexts, \texttt{github}~\cite{gh_context} and \texttt{secrets}~\cite{secrets_context} are the most used ones. 

The \texttt{github} context contains the event that triggered the workflow run plus some further information.
It can be involved in the workflow execution to provide information like the actor who triggered the workflow or the body text of a newly created issue.
The \texttt{secrets} context contains the names and values of secrets that are available to a workflow run.
Secrets can be used to manage confidential information, like API keys and passwords. For example, the \texttt{GITHUB\_TOKEN} is automatically included in the \texttt{secrets} context.
Finally, GHA offers the possibility to make workflows reusable \cite{reuse_wf}. This mechanism enables anyone with access to the repository and the reusable workflow to call it from another workflow. Workflow reuse also promotes best practices by helping developers to use workflows that are well designed, have already been tested, and have been proven to be effective. Also, reusable workflows enable the definition of organization-wide libraries of workflows that can be used to speed up the creation of CI/CD pipelines.

\subsection{Security Issues of GHA workflows}
\label{subsec:workflow-vulnerabilities}
The introduction of workflows enabled code repositories to become an integral part of the CI/CD pipeline. In particular, GHA workflows can operate on the code repository by programmatically adding, removing, and modifying its content. 
Those capabilities, however, can directly affect the confidentiality, integrity, and availability of the software and the associated metadata information.

In detail, an attacker can leverage security weaknesses and misconfigurations in the definition and execution of a GHA workflow.
As some elements of the workflow can be manipulated by the actor triggering the action, a malicious actor can craft specific inputs to cause unexpected execution flows in the workflow.

For instance, context elements are susceptible to several security weaknesses and misuse, as stated in the official documentation \cite{gh_context}.
On one hand, the \texttt{secrets} context can contain sensitive information (e.g., a token or a key) that is encrypted using the default key and can be safely manipulated inside a workflow environment. On the other hand, a poorly configured workflow may grant direct access to a secret, thereby allowing an attacker either to send it to unintended hosts or explicitly print it to the log output \cite{gha_secrets}.

As for secrets, direct access to variables of the \texttt{github} context can lead to command injection attacks.
In detail, the \texttt{github} context allows storing information that directly depends on the user's input (e.g., the body of an event). Those data can be manipulated and executed inside by a Runner using the \texttt{run} job. Unfortunately, a poorly-configured workflow can allow attackers to inject a crafted input to trigger its direct execution inside the Runner, thereby compromising the repository or the execution environment. 

\begin{figure}
    \centering
    \includegraphics[width=\columnwidth]{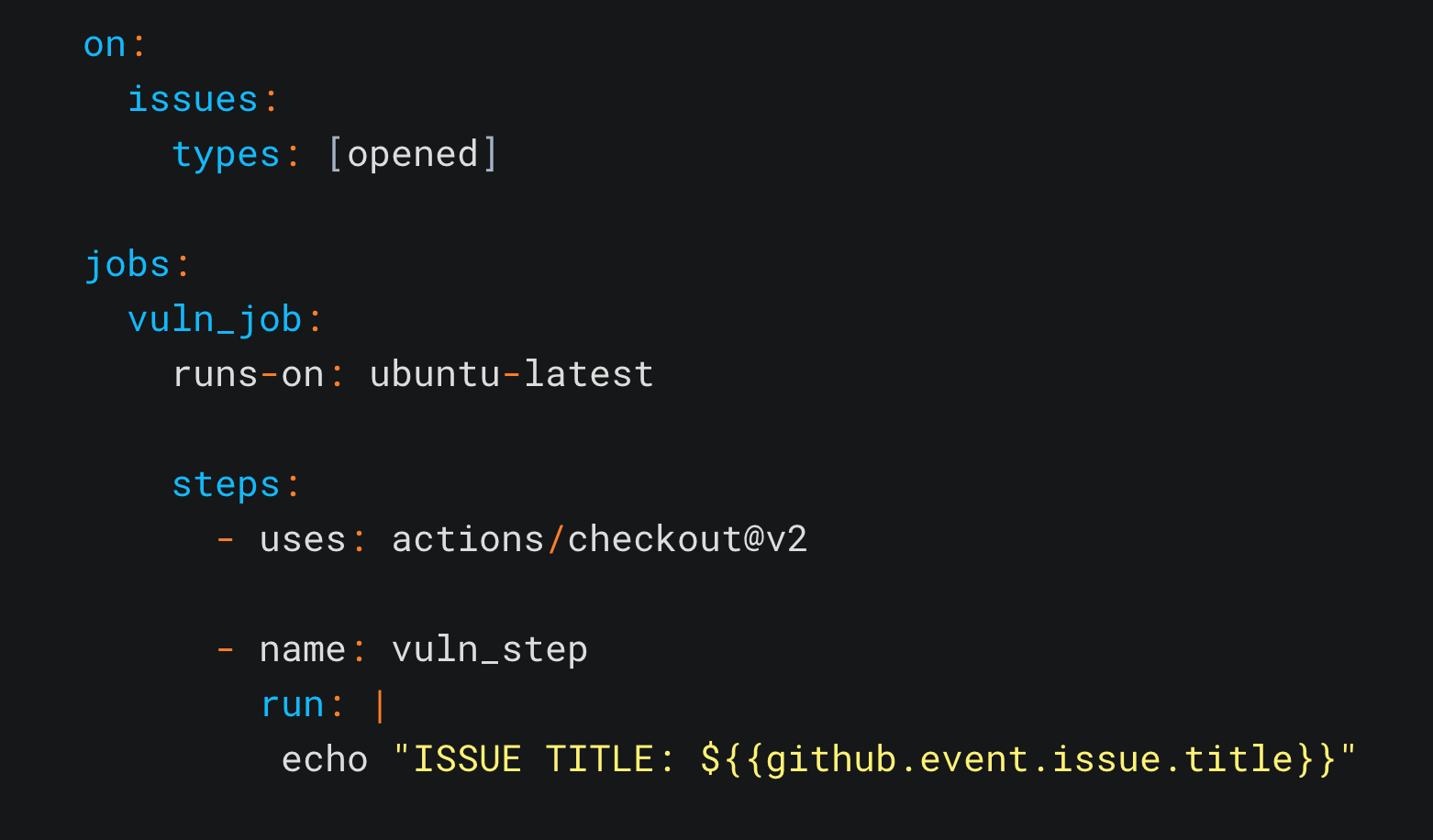}
    \caption{Example of Github Workflow vulnerable to command injection attacks.}
    \label{fig:vuln_wf}
\end{figure}

An example of vulnerable scenario is depicted in Figure~\ref{fig:vuln_wf}.
In this scenario, the \texttt{issue.title} element is directly executed in the run step. If an attacker can inject in the title of the issue the string: 
\texttt{\small{New malicious issue title” \&\& bash -i >\& /dev/tcp/0.tcp.eu.\-ngrok.\-io/14872 0>\&1 \&\& echo ”}}, she can obtain a reverse shell on the remote machine hosting the Runner.

In addition to the security weaknesses of workflow elements, two other features of GHA have direct impact on the security of workflows, i.e., \textsl{permissions} and \textsl{reusable workflows}.

In the first case, permissions can affect the capability of a successful attack on a repository~\cite{gha_sec_perms}. To this aim, developers should enforce the least privilege principle by assigning the least set of permissions tokens and workflows, and differentiate - when possible - the set of permissions granted to each job.

Finally, reusable workflows allow repository owners to import both workflows belonging to their organization and workflows developed by third parties that are publicly available \cite{gha_tpwf}.
In this latter case, the caller workflow does not control the imported one (the callee). The introduction of unmonitored workflows increases the attack surface since the callee may include potential weaknesses and misconfigurations, especially if the imported version is not updated.
Also, attackers can manipulate reusable workflows by exploiting tag-reuse attacks to trick the caller workflow into importing a different version of the callee, which has been maliciously re-tagged.
To prevent such issues, the best practice suggests selecting the workflow version using a commit hash (e.g., \texttt{workflow\_name@cdd...08c})~\cite{gha_tpwf}.

\section{Security Assessment Methodology}
\label{sec:contribution}
We propose a novel methodology composed of two phases, namely \emph{Workflow Collection} and \emph{Workflow Security Evaluation} in order to evaluate the security of GHA in a software supply chain.
The Workflow Collection phase is devoted to analyze the SSC and extract a model that includes all the involved CRs and - for each repository - the set of GHA workflows.
The Workflow Security Evaluation phase performs a security analysis of each GHA workflow to assess five security categories extracted from the analysis of the official documentation \cite{gh_actions} and the security hardening guidelines \cite{gha_hardening}.
The result of the analysis is a report containing a list of security issues affecting the workflows used in the SSC, classified according to their exploitability.

\subsection{Workflow Collection}
\label{subsec:workflows_collection}

The first part of the methodology takes as input a folder containing the software under test (hereafter, SUT).
The procedure parses the code and configuration files of the project to recursively identify software dependencies, code repositories, and distribution networks composing the Software Supply Chain of the SUT.
From such a piece of knowledge, the methodology builds a direct graph structure where nodes include any code repository involved in the SSC, while edges represent the relationships among them.

Then, the procedure focuses on each CR to extract its GHA workflows (if any).
More specifically, the methodology relies on GitHub API to search and retrieve the YAML files of the available workflows. The set of workflows is then linked to the corresponding CR node in the graph.

\subsection{Workflow Security Evaluation}
\label{subsec:evaluation}

The Workflow Security Evaluation phase scans the set of GHA workflows to detect security vulnerabilities and misconfigurations that can affect code repositories in the SSC. 

The detection logic of this phase is based on the evaluation of a set of \emph{security categories} that are mapped in constraints - i.e., security requirements that workflows must comply with - and \emph{security checks} - i.e., technical controls to assess the enforcement of constraints.
The list of categories, constraints, and checks applicable to GHA workflows derive from a manual review of the GitHub guidelines for workflow hardening \cite{gha_hardening} and from the official GHA documentation \cite{gh_actions}. Table \ref{tab:gha_requirements} reports the results of our analysis.

In particular, we identified four categories from the analysis of GH guidelines concerning the security of GHA workflows, namely \emph{Confidential Data Disclosure}, \emph{Command Injection}, \emph{Third-party Workflows}, and \emph{Workflow Permissions}. Also, we extended those categories with \emph{Triggering Events} as they represent the entry point for workflows.

\begin{table*}[]
    \centering
    \resizebox{2\columnwidth}{!}{\begin{tabular}{|l|l|l|}
        \hline
        \textbf{Security Categories} & \textbf{Security Constraints} & \textbf{Security Checks}\\
        \hline
        Confidential Data Disclosure & - Registering generated values as secrets. & (SC-1) Check the use of the \texttt{secrets} context outside environment.\\
            & - Registering modified values as secrets. & (SC-2) Check the use of \texttt{secrets} context for generating new data.\\
        \hline
        Command Injection & - Do not directly use github context in scripts. & (SC-3) Check the use of \texttt{github} subcontexts inside run steps. \\
        \hline
        Third-Party Workflows & - Audit the use of branch names, emails, and external inputs. & (SC-4) Verify the version of reusable workflows.\\
            & - Audit information passed to TP actions. & (SC-5) Verify the use of the pinning Commit Tag.\\
            & - Keep third-party workflows up-to-date. & \\
        \hline
        Workflow Permissions & - Specify permissions to avoid uncontrolled access. & (SC-6) Check if the workflows enforce the least privilege principle.\\
        \hline
        Triggering Events & - Audit the kind of events that trigger the workflow. & (SC-7) Verify which is the exploitability score of the event. \\
            & - Audit the filters applied to triggering events. &  \\
        \hline
    \end{tabular}}
    \caption{List of security categories, requirements, and checks of the Workflow Security Evaluation phase.}
    \label{tab:gha_requirements}
\end{table*}

The evaluation phase parses each GHA workflow to \emph{i}) execute the security checks detailed in Table \ref{tab:gha_requirements} to detect security vulnerabilities and misconfigurations, and \emph{ii)} evaluate the exploitability of the findings by identifying the events and preconditions that allow triggering the corresponding security issue.

\begin{figure}
    \centering
    \includegraphics[width=\columnwidth]{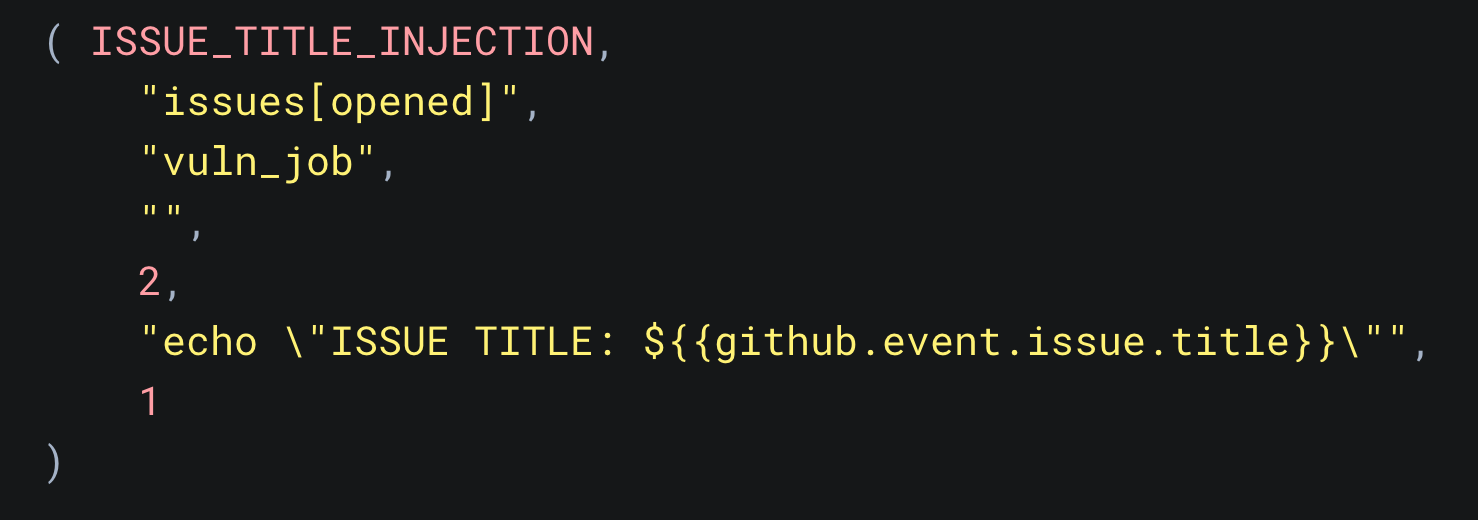}
    \caption{Example of a COMMAND\_INJECTION security issue detected in the workflow of Figure \ref{fig:vuln_wf}.}
    \label{fig:tuple_ex}
\end{figure}

The result of the evaluation is a set of tuples containing the detected security issue, some information related to the issue (i.e., the name of the job and the affected step, the position of the issue, and the print of the affected line), the event triggering the issue (i.e., the entry point), and an evaluation on its exploitability. If an issue belongs to a job triggered by multiple events, the evaluation will contain a separate tuple for each event to enable the filtering of specific results.

\begin{lstlisting}[language=json, caption=Structure of a security issue identified by the Workflow Security Evaluation phase., captionpos=b, frame=lines, label=lst:tuple_template]
       ( security_issue_type,
         triggering_event,
         job_name,
         step_name,
         step_position,
         issue_line,
         exploitability_score )
\end{lstlisting}

Listing \ref{lst:tuple_template} provides the structure of a tuple, while Figure~\ref{fig:tuple_ex} shows an example of a tuple for the scenario depicted in Figure~\ref{fig:vuln_wf}.

The rest of this section will detail the evaluation of the exploitability of events and the type of security issues supported by the methodology.

\subsubsection{Execution of Security Checks.}

The execution of the security checks listed in Table \ref{tab:gha_requirements} enables the identification of a set of security issues targeting the workflow under test.
The methodology labels an identified security issue into two different groups, i.e.:
\begin{description}
    \item[Vulnerability.] A flaw in the workflow that is directly exploitable. For example, a misuse of the \texttt{git\-hub.\-con\-text} API enables attackers to execute command injection attacks.
    \item[Misconfiguration.] A configuration error that makes the environment, the CR, or the SSC vulnerable. For example, the improper configuration of \texttt{tag} variable inside a workflow may lead to the import of an outdated third-party workflow.
\end{description}

In detail, the security issues belonging to the Confidential Data Disclosure and the Command Injection categories can be identified using pattern verification techniques.

Secrets can be used in workflows through the \texttt{secrets} context~\cite{secrets_context}.
A particular secret is accessed declaring the context and then its name.
Secrets are expected to be accessed inside of an environment.
Indeed, when they are accessed from outside, secrets can be exfiltrated by an attacker (e.g., printed in a log or sent to a remote host) that has access to the workflow.
Then, the security checks for Secrets consist of assessing if secrets are used in the environment part of the workflows (\texttt{SC-1}) and, if it is not the case, if they are manipulated in order to generate confidential data (\texttt{SC-2}).
Indeed, \texttt{SC-2} allows assessing when a secret is involved in a computation and then exposed to potential exfiltration.

Similarly, the methodology can verify the presence of command injection attacks in workflows by evaluating the usage of \texttt{github} context~\cite{gh_context} in run steps.
In detail, the syntax \texttt{github.*} is used to access variables contained in the GitHub Context.
This context contains many subcontexts linked to different aspects of the workflows.
For example, the \texttt{github.event} context contains all the information of the triggering event.
Some parts of the \texttt{github} context can be manipulated externally by an actor.
Hence, an attacker can inject a command into a vulnerable instruction through this context.
For example, in the scenario of Figure~\ref{fig:vuln_wf}, the attacker is able to exploit the run step through the issue title contained in \texttt{github.event.issue.title}.
For this reason, the methodology evaluates the use of \texttt{github} subcontexts that can be exploited by an attacker to affect the workflow run (\texttt{SC-3}).

The command injection issues is further distinguished between \textit{conditional} and \textit{unconditional}. The first case consists of command injection attacks in a branch of a conditional statement, thereby requiring the attacker to trigger the appropriate branch to exploit the vulnerability. In the unconditional case, the vulnerable statement 
will be executed directly.

To evaluate the security checks belonging to the Third-party Workflows category, the methodology relies on the GitHub API to obtain the latest version of the specific TP workflow, and compare it with the version requested by the workflow under test.
If the used TP workflow is out-of-date (\texttt{SC-4}) or the workflow under test does not pin a specific commit hash (i.e., using the \texttt{<wf\_name>@<commit\_hash>} syntax) (\texttt{SC-5)}, the methodology marks the issue as a security misconfiguration.

Finally, the methodology evaluates the workflow under test to check if the permissions declared in the workflow follow the principle of least privilege. Following the least privilege principle, the methodology checks if the required permission matches the capabilities required by the workflow (\texttt{SC-6}).

\subsubsection{Exploitability Score}

The only way for an attacker to execute a workflow and, thus, exploit a security issue is by triggering one or more events defined in the workflow. 
To this aim, the methodology evaluates the potential entry points by assessing the configuration of workflow events and the type of conditions that enables their activation (\texttt{SC-7}).

Events are influenced by filters \cite{gha_filters} that can be applied to specialize the triggering actions. For example, an \textit{Issues} event can have up to 16 activity types (e.g., opened, deleted, labeled, \ldots). To this aim, the use of filters directly affects the exploitability of a workflow.

The methodology classifies events using three security levels based on the complexity an attacker has to face to stimulate a particular event.
We consider the complexity as the number and type of preparatory activities to get the required assets (e.g., privileges, knowledge) to trigger an event.

\begin{description}
    \item[Restricted (1)] The attacker needs to achieve a multi-stage attack to be able to trigger the specific workflow event.
    The attacker can succeed only with the help, whether intentionally or not, of one of the owners of the repositories.
    For example, an attacker cannot trigger a push event unless a repository owner grants him the maintainer status.
    \item[Supervised (2)] The attacker must match particular conditions in order to trigger the event.
    For example, a pull request can be triggered, but the maintainer has to accept it and consequently start the workflow.
    \item[Unsupervised (3)] The attacker does not need any granted permission by the repository owners. He can trigger the workflow at any time by means of external action.
    For example, an issue opening event can be triggered by any GitHub user with access to the repository.
\end{description}

\section{Prototype Implementation}
\label{sec:implementation}

\begin{figure*}
     \centering
     \includegraphics[width=\textwidth]{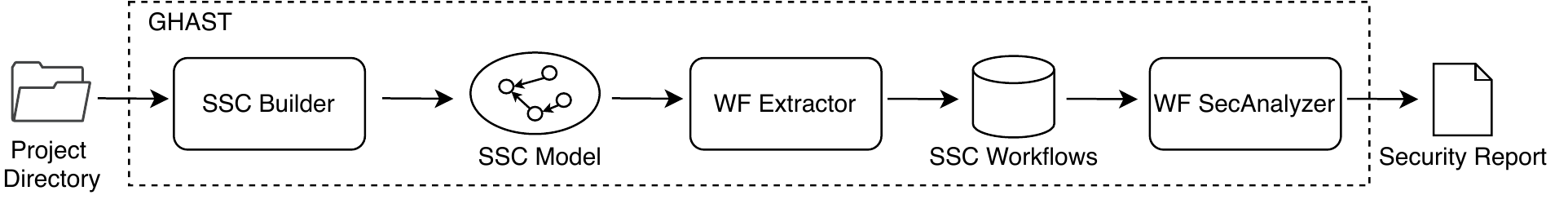}
     \caption{The \toolname~architecture.}
     \label{fig:wf_plugin}
 \end{figure*}

We implemented the proposed methodology in a command-line tool - the GitHub Actions Security Tester (\toolname)~- composed of three modules, as depicted in Figure~\ref{fig:wf_plugin}. The source code of \toolname~is publicly available\footnote{https://github.com/Mobile-IoT-Security-Lab/GHAST}.

The first module (\emph{SSC Builder}) extracts the model of the SSC starting from a folder containing the source code of SUT.
The second one (\emph{WF Extractor}) extracts the list of code repositories from the SSC model.
It extracts the set of GHA workflows from these collected repositories.
The last module (\emph{WF SecAnalyzer}) executes the workflow security evaluation described in Section \ref{subsec:evaluation} to produce the final security report.

For the \textit{SSC builder}, we rely on a state-of-the-art research tool for the rebuilding and modeling of software supply chains \cite{Benedetti2022AliceEvaluation}.
In detail, the module requires the folder containing the SUT as input, then it recursively identifies the SSC assets, including code repositories. The output model is a direct graph structure stored in a Neo4J Graph database~\cite{neo4j}. 

The \textit{WF Extractor} queries the SSC model to extract the set of repositories that uses GHA workflows. Then, the module scans each repository using the GitHub API and downloads all the workflow YAML files contained in the corresponding \texttt{.git\-hub/\-workflows} folder. The result of this operation is the collection of workflows used in the SSC.

Finally, the \textit{WF SecAnalyzer} evaluates each workflow by applying the security checks listed in Table \ref{tab:gha_requirements}. To do so, the module parses each YAML file to extract the following GHA elements:
\begin{itemize}
    \item \textit{triggering-events}, for the computation of the exploitability score (\texttt{SC-7}).
    These elements also contain event filters that are included in the score computation.
    \item \textit{Runs}, for the evaluation of command injection and confidential data disclosure issues (\texttt{SC-1}, \texttt{SC-2}, \texttt{SC-3}).
    \item \textit{TP-workflows}, for the evaluation of commit pinning and the workflow versions (\texttt{SC-4}, \texttt{SC-5}).
    \item \textit{Permissions}, for the evaluation of permissions both at workflow level and at job level (\texttt{SC-6}).
\end{itemize}

\begin{lstlisting}[language=json, caption=Extract of a sample security report produced by \toolname., captionpos=b, label=lst:ex_output]
"<work_test>": 
{
  "events": ["issues",3],
  "issues": [
    ["MISCONF_PERM_GLOBAL", "<job_A>",...],
    ["OUTDATED_WF", "<job_A>",...],
    ["CI_ACTOR","<job_B>",...],
    ["OUTDATED_WF", "<job_A>",...],
    ["OUTDATED_WF", "<job_B>",...]
  ]
}
\end{lstlisting}

The output of the evaluation is a security report in JSON format that contains - for each repository (and workflow) of the SSC - the set of security issues identified by the tool. Listing~\ref{lst:ex_output} shows an extract of an output file. In the example, the tool identified five issues in the workflow \texttt{work\_test}, associated with the event \texttt{issues}, i.e., a command injection exploiting the actor username, a misconfiguration of the permissions, and three outdated third-party workflows. The exploitability score of the event triggering those finding is 3, i.e., unsupervised.

\section{Experimental Evaluation}
\label{sec:results}

We conducted an experimental campaign to test the applicability and efficacy of \toolname~in the wild on software repositories available on GitHub.
In detail, we ran the tool against 50 public repositories randomly taken from the top 100 Python-based projects on GitHub in July 2022. The repositories are actively maintained and used by the community (e.g., the dataset has an average number of stargazers per repository above 20k). Table \ref{tab:repos} details the dataset's characteristics regarding the number of stargazers, forks, contributors, open issues, and commits.

\begin{table}[tbh]
    \centering
    \begin{tabular}{|c|c|c|c|}
    \hline
        & \textbf{MIN} & \textbf{MAX} & \textbf{AVG} \\
        \hline
       \textbf{Stargazers} & 4,365 & 191,386 & 22,227  \\
       \hline
        \textbf{Forks} & 307 & 36,334 & 4,708 \\
        \hline
        \textbf{Contributors} & 3 & 30 & 25 \\
        \hline
        \textbf{Issues} & 1 & 1,707 & 232 \\
        \hline
        \textbf{Commits} & 40 & 52,887 & 4,729  \\
        \hline
        
    \end{tabular}
    \caption{Number of stargazers, forks, contributions, issues and commits (min, max, and avg values) of the dataset.}
    \label{tab:repos}
\end{table}

The tool extracts the SSC for each repository, identifies the CRs compatible with GHA workflows, and processes the security report. The experiments were hosted on a virtual machine running Ubuntu 20.04 with 8 processors and 32GB RAM.

\subsection{Experimental Results}

\toolname~was able to reconstruct the SSC of the 50 projects obtaining 646 unique CRs in approximately 14 hours. Then, the tool extracted 
131,168 GHA workflows and executed the security checks.

The analysis allowed for the identification of 24,905 security issues, i.e., 20 security vulnerabilities and 24,885 misconfigurations.
All the findings have been directly reported to the maintainers of the repositories.

\begin{figure*}
    \begin{minipage}[b]{0.5\textwidth}
    \includegraphics[width=\columnwidth]{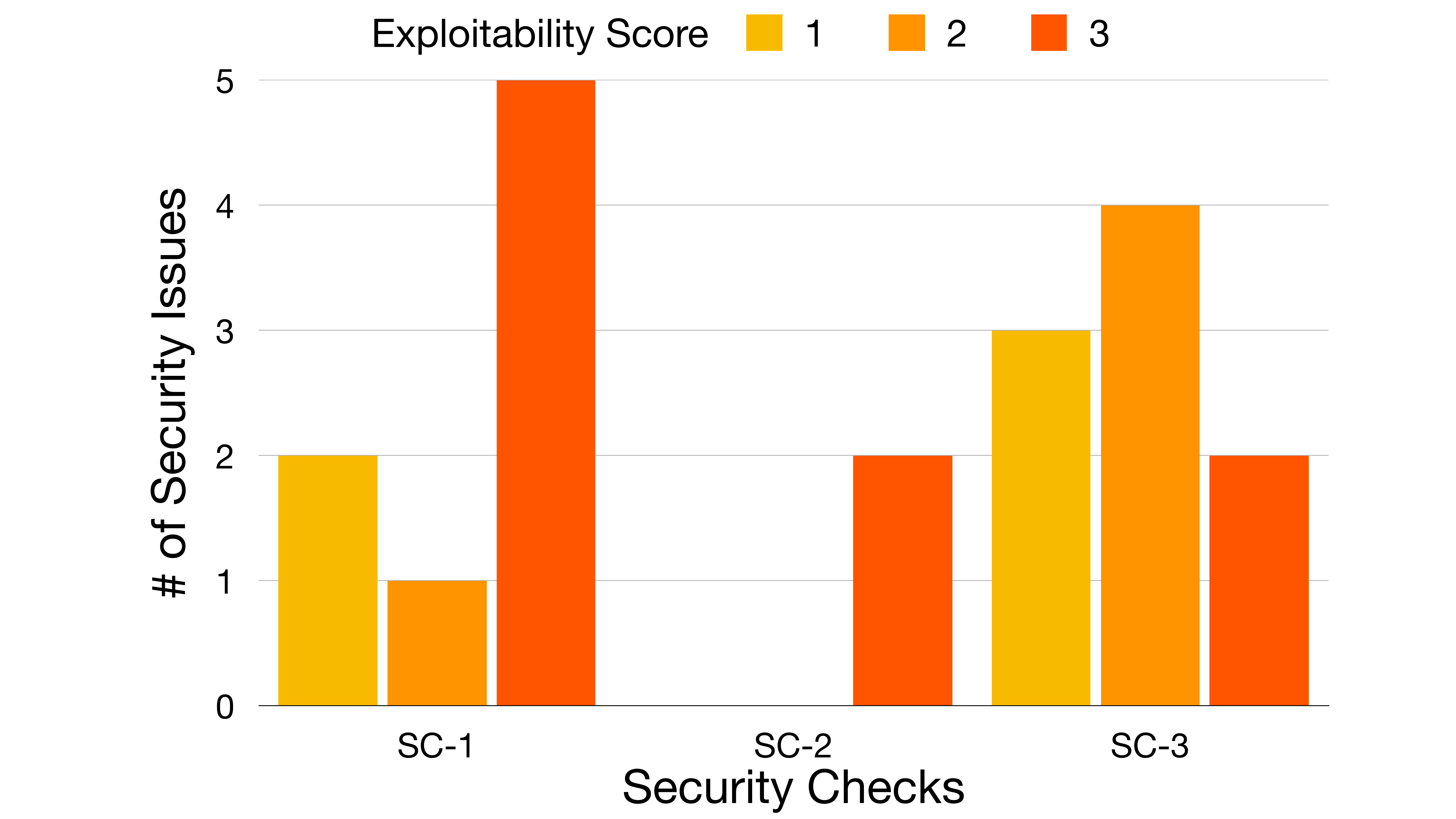}\\
    \subcaption{Vulnerability Issues identified.}
    \label{fig:vuln_issues}
    \end{minipage}%
    \begin{minipage}[b]{0.5\textwidth}
    \includegraphics[width=\columnwidth]{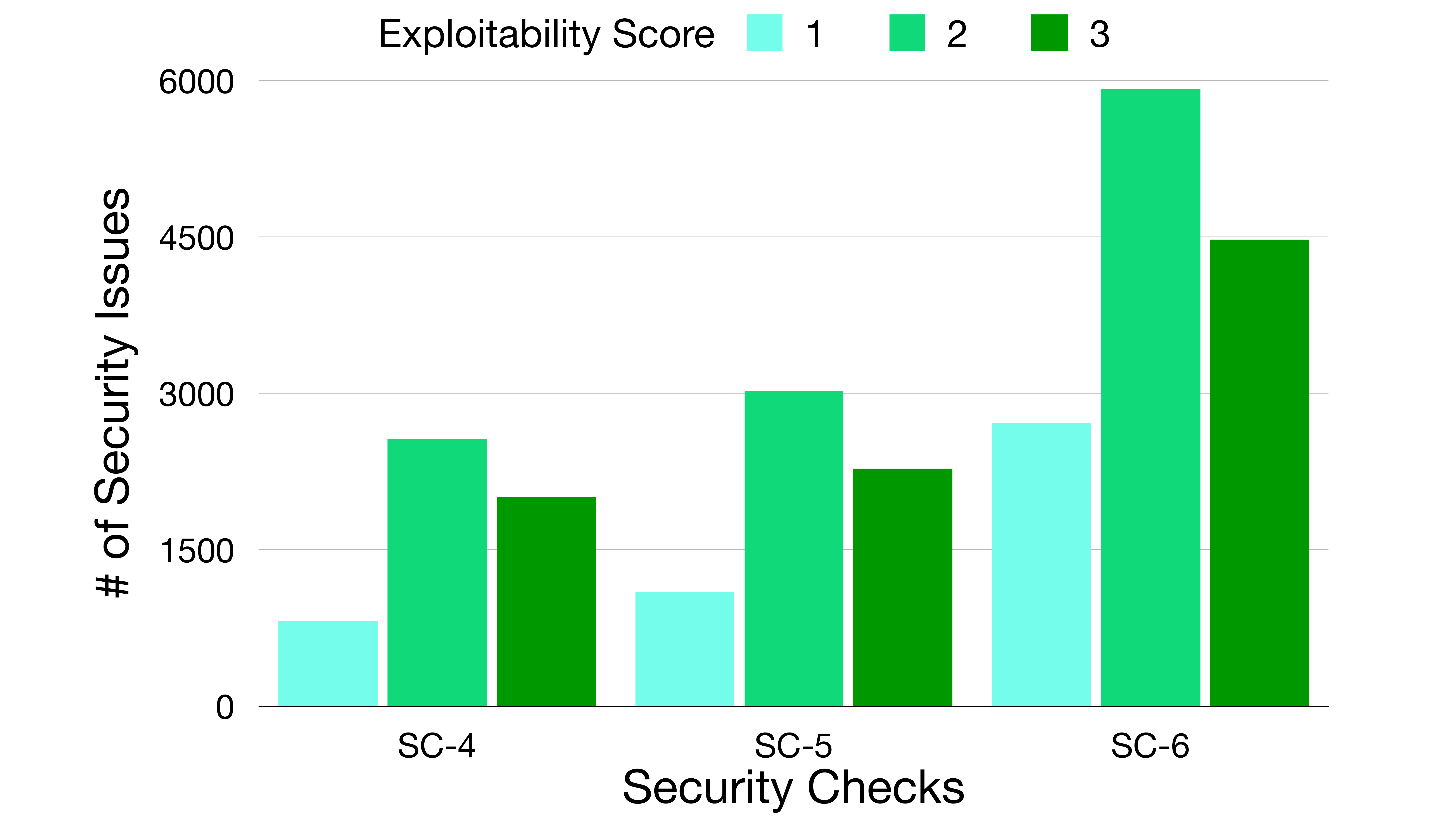}\\
    \subcaption{Misconfiguration Issues identified.}
    \end{minipage}%
    \caption{Issues for different exploitability scores.}
\end{figure*}

\paragraph*{Security Vulnerabilities.}
Figure~\ref{fig:vuln_issues} reports the distribution of security vulnerabilities identified using \texttt{SC-1}, \texttt{SC-2}, and \texttt{SC-3} according to their exploitability score.

The security checks on Secrets \texttt{SC-1} and \texttt{SC-2} reported 10 jobs that are not using the \texttt{secrets} context inside an environment. Such behavior enables jobs to expose those secrets, e.g., by printing in a log output or sending them to an external host. Also, it is worth noticing that most of these vulnerabilities (i.e., 7 out of 10) can be triggered using unsupervised events.

\toolname~also reported 9 workflows vulnerable to command injection attacks (\texttt{SC-3}), of which two do not need any granted permission by the repository owner to be exploited (score equals to 3). Also, 3 out of 9 are \emph{unconditional} command injections, i.e., that the affected step is not included in any conditional branch, thereby easing the exploitation process.

\paragraph*{Misconfigurations.}
For \texttt{SC-4} and \texttt{SC-5}, we considered both checks passed when the workflow uses the commit hash of the latest version available of the third-party workflow. In particular, we found that this condition has never been verified on our sample set, meaning that even if a workflow uses the \texttt{latest} tag, it is still vulnerable to tag-reuse attacks. 
The experimental campaign allowed the identification of 5,384 steps that do not use the latest version of a third-party workflow. Also, the evaluation of \texttt{SC-5} resulted in 6,388 steps that do not pin a specific version of a third-party workflow. 

\begin{figure}
    \includegraphics[width=0.8\columnwidth]{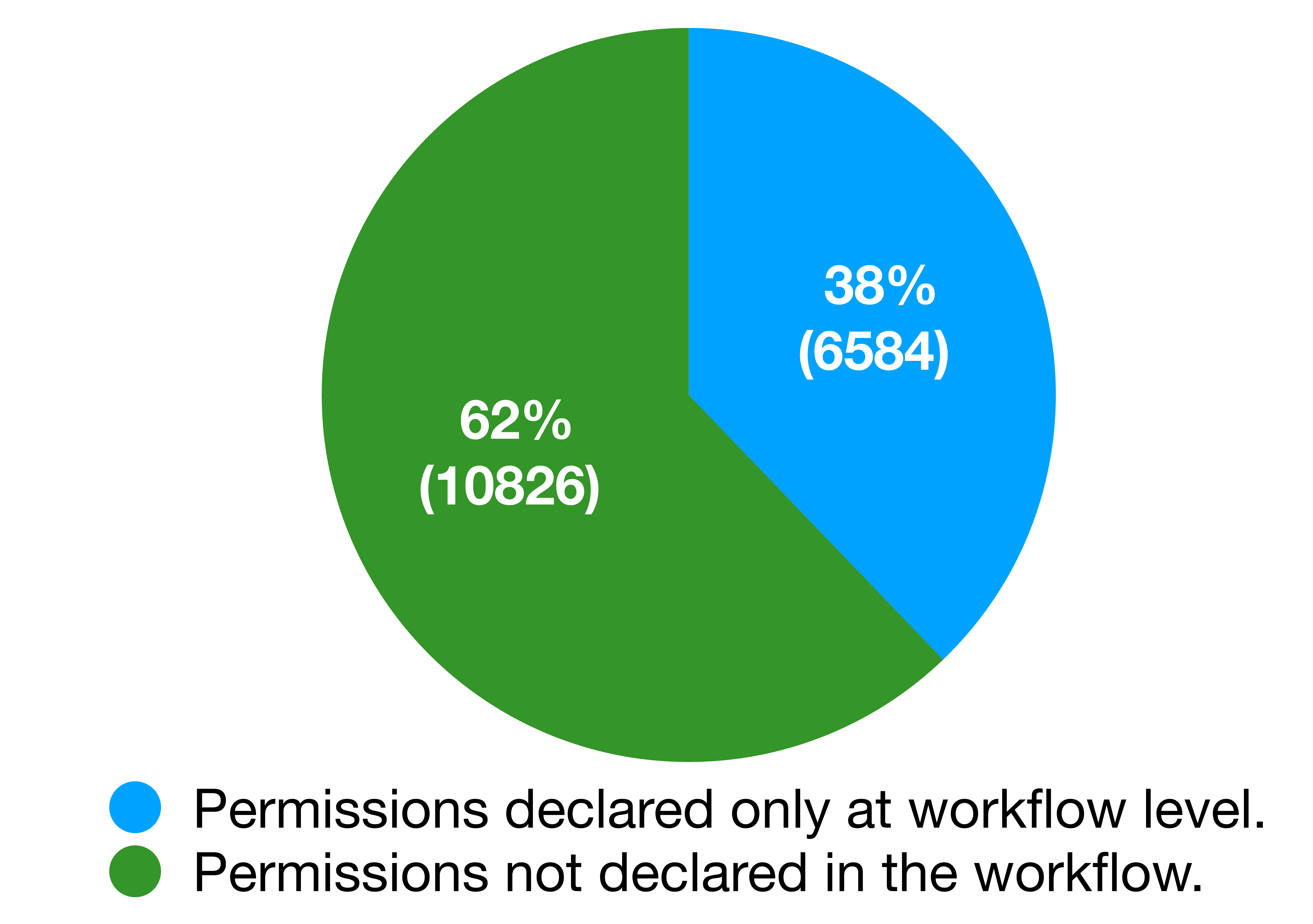}\\
    \caption{Distribution of Misconfigured Permissions (SC-6).}
    \label{fig:percentages}
\end{figure}

Finally, \texttt{SC-6} reported several misconfigurations in the definition of workflow permissions. In detail, Figure~\ref{fig:percentages} shows that 38\% of workflows declared the permissions granted for their execution at the workflow level instead of for each job (as suggested by the security guidelines~\cite{gha_hardening}). Furthermore, 62\% of workflows do not declare any specific permission, thereby allowing the execution of the workflow with the default permissions assigned to the \texttt{GITHUB\_TOKEN}.

\subsection{Manual Validation of the Vulnerabilities}

We manually revised the 20 identified vulnerabilities in order to verify the presence of false positives.
To conduct this analysis, we downloaded the affected repositories locally and instantiated a local GHA Runner for the corresponding workflows to reproduce the exact conditions to test the exploitability of the attack. 
The result of manual analysis confirmed that all the vulnerabilities identified by \toolname~are true positives.

Listing~\ref{lst:vuln_pull_req} shows an anonymized GHA workflow containing a command injection issue triggered by a pull request event reported by \toolname. The workflow belongs to a repository with more than 10,000 stars, 500 forks, and 3,000 commits.

\begin{lstlisting}[language=json, caption=Excerpt (anonymized) of a workflow vulnerable to a command injection attack., captionpos=b, label=lst:vuln_pull_req]
name: Pull Request Validation
on:
  pull_request:
    types: [opened, synchronize, reopened, edited]
jobs:
  <anon-job-name>:
    name: -------------
    runs-on: ubuntu-latest
    steps:
    - name: <anon-step-name>
      uses: actions/checkout@v2
      with:
        ref: ${{github.event.pull_request.head.sha}}
        fetch-depth: 0
    ...
    - name: <anon-step-name>
      run: |
        cat << EOF | egrep -qsi '^disable-check:.*\<commit-count\>'
        ${{github.event.pull_request.body}}
        EOF
\end{lstlisting}

In such a workflow, it is possible to replicate a similar attack that affects the workflow in Figure~\ref{fig:vuln_wf} to initiate a reverse shell and access the Runner executing the workflow.

\begin{figure}[h]
    \centering
    \includegraphics[width=\columnwidth]{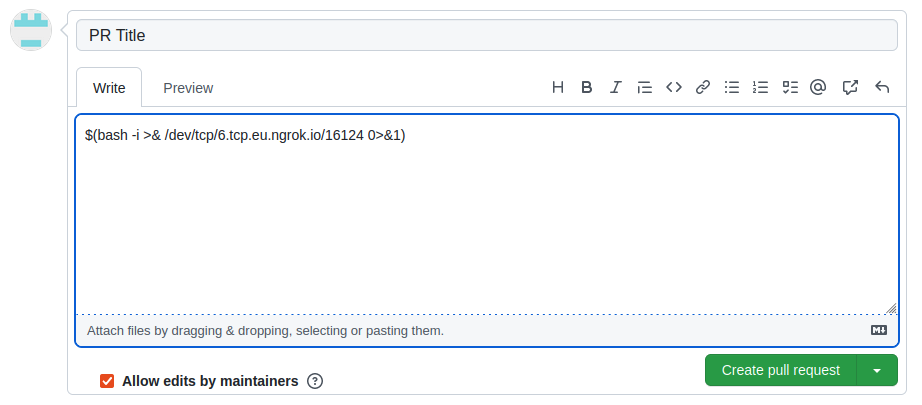}
    \caption{Creation of a malicious pull request on the target repository.}
    \label{fig:pr_expl}
\end{figure}

In detail, we crafted a payload to exploit the interpolation of the \texttt{github.event.pull\-\_request.body} variable in the workflow, and we submitted it as a new pull request in the target repository, as shown in Figure~\ref{fig:pr_expl}. 
If a maintainer accepts the pull request, she will trigger the vulnerable workflow.
When the GHA reaches the run step of Listing~\ref{lst:vuln_pull_req}, the Runner executes rows 17-20 using bash.

\begin{figure}[h]
    \centering
    \includegraphics[width=\columnwidth]{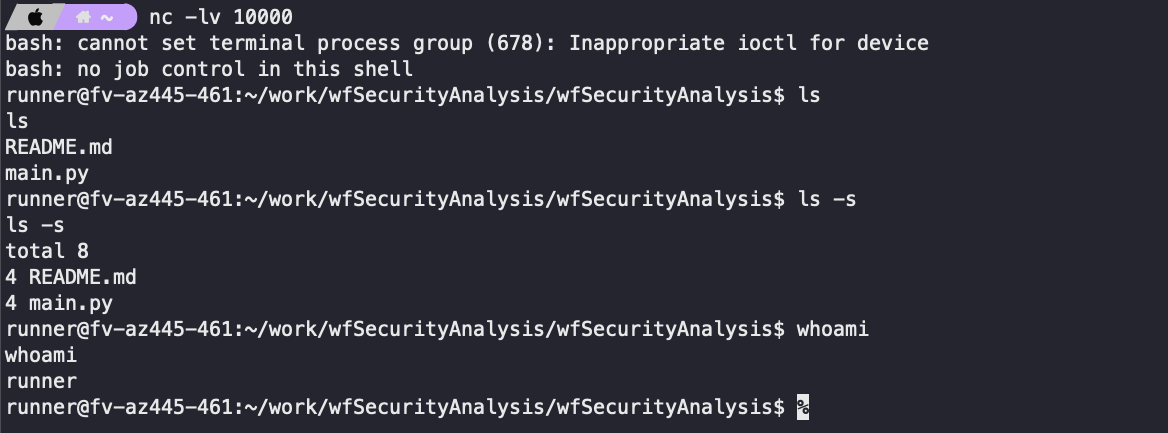}
    \caption{Attacker terminal with the reverse shell to the affected Runner.}
    \label{fig:runner_expl}
\end{figure}

In this case, the run step contains an \texttt{here documents\- re\-di\-re\-ction}~\cite{heredoc} that allows command substitution~\cite{cmd_sub}. 
Hence, the GHA Runner executes the malicious payload from the pull request and opens a TCP connection towards a remote server on port 16147.
Figure~\ref{fig:runner_expl} shows the listener active on local port 10000 mapped to the remote server, where the reverse shell is open. At this point, the attacker has complete access to the Runner instance, e.g., she can execute any command with Runner's privileges.

\section{Related Work}
\label{sec:related_work}

To the best of our knowledge, the research work targeting the evaluation of GHA in the context of the software supply chain is limited.
On the one hand, several organizations like ENISA \cite{enisa}, NIST \cite{nist}, and OWASP \cite{owasp} discussed the security issues in the Supply Chain, giving a focus on SSCs and the impact of using insecure third-party software in the development pipeline. Also, several technical reports investigated attacks targeting SSCs of popular software like Solarwind \cite{herr_broken_2021} and Log4J \cite{log4shell}. 

On the other hand, the scientific community mainly investigated the problem of ensuring the integrity of the final software in SSCs that include different actors (e.g., code repository platforms, software library binaries, distribution networks).
Several works \cite{repr_builds, npm_repr} propose the use of reproducible builds to create an independently-verifiable path from source to binary code.
To cope with the integrity problem in the last part of the development process, Vu et al. in~\cite{lastpymile} show how the software can be compromised between the production of source code and the building process and how it is possible to mitigate this security problem.
Considering the entire CI/CD pipeline, the in-toto framework~\cite{in-toto} proposes holistic integrity enforcement of a software supply chain. It gathers cryptographically verifiable information about the chain to accomplish its objectives.

To the best of our knowledge, only two open source projects explicitly targetting the evaluation of code repositories: GitGat~\cite{gitgat} and GitHub Workflow Auditor ~\cite{gh-tinder}. 
The first one is a project released in June 2022 that takes advantage of the Open Policy Agent (OPA) ~\cite{opa} to evaluate security policies for GitHub's organization, repositories, and user accounts.
GitHub Workflow Auditor is a command line tool released by Tinder in July 2022 for the security assessment of workflows. The tool scans a specific organization, user, or repository to detect potential security issues in secrets usage and external inputs. 
Unlike \toolname, GitHub Workflow Auditor can neither reconstruct the SSC of a project nor extract all the workflows associated with different CRs to detect security issues. Also, the tool does not provide any security evaluation of Third-party Workflows, Workflow Permissions, and Triggering Events.
Also, it is worth noticing that both tools were released in Q2 2022, thereby suggesting a growing interest in the topic.

Finally, Koishybayev et al. \cite{usenix_article} studied the security of GitHub CI workflows in parallel with our research. In detail, their work identified four security properties (permissions, privileges, code controls, and secrets) affecting workflows in GitHub CI and other VCS platforms. Also, they released a PoC tool called GWChecker to assist in analyzing GHA workflows.
As for the other works on code repositories, their research focuses on single workflows, and neither takes into account nor models the dependencies among GHA workflows, thereby lacking an evaluation of the workflows on the entire software supply chain of the SUT. In addition, the authors do not consider events and their exploitability to evaluate the impact of vulnerabilities and security misconfigurations of GHA workflows.

\section{Limitations and Discussion}
\label{limitations_discussion}

\paragraph*{Limitations.}
The proposed methodology obtained promising results during the experimental evaluation and allowed us to assess the applicability and efficacy of GHAST. In addition, the manual validation of the identified vulnerabilities confirms the reliability of the approach.

Still, our solution suffers from some limitations. First, the security evaluation methodology has inherited limits. 
Although some security controls (e.g., SC-4 and SC-5) are based on the definition syntax of GHA workflows, having an intrinsically low rate of false positives.
Other SCs (e.g., SC-1 and SC-2) are based on regular expressions and pattern matching techniques that cannot keep into account other information, like, e.g., the workflow execution context. As a consequence, this could, in principle, lead to potential false positives.

In our work, we tried to reduce the rate of false positives by thoroughly reviewing the GHA documentation and real-world GHA workflows.
This approach allowed us to catch some potential corner cases and forge a heuristic to evaluate the SCs. However, it is still possible that GHAST does not identify deviant cases; therefore, we argue that an assessment in the wild may lead to false positives.

Moreover, the evaluation of the GHA workflows depends on correctly identifying all the software repositories involved in the SSC and the associated GHA workflows. 
To this aim, an error in the parsing process of the SUT can lead to not identifying a subset of CRs and the related workflows, thereby leading to potential false negatives.
In this respect, GHAST is based on the Sunset Framework \cite{Benedetti2022AliceEvaluation} and shares the same parsing limitations.

\paragraph*{Vulnerability Disclosure and Security Implications.}
We disclosed all the identified vulnerabilities to the owners of the affected repositories via email. Each email contained a description of the vulnerability, the potential impact of its exploitation, and the GHAST report.
At the time of writing, 6 out of 20 repository owners acknowledged our notifications within a period of 1 month after the disclosure.

As with any security assessment tool, attackers can use GHAST to discover vulnerable projects. 
When releasing a tool, there is a fundamental trade-off between helping repository owners versus facilitating the attackers.

Given the increased attention to software supply chain security and the existence of similar publicly available tools (e.g., \cite{usenix_article}~\cite{gitgat}, and ~\cite{gh-tinder}), we feel that the benefits to repository owners for publicly releasing GHAST outweigh the harms. 
Attackers have the resources and capabilities to replicate GHAST, whereas many repository owners do not.
Also, GHAST does not explicitly report how to exploit security misconfigurations. 
Finally, publicly releasing GHAST will also encourage further research and would help repository maintainers (e.g., tools to patch vulnerable workflows automatically).

\section{Conclusion}
\label{sec:conclusion}
We investigated the security issues affecting GHA workflows to understand their security impact on the software supply chain.
We produced an analysis of GHA aspects impacting the security of repositories.
Then, we leveraged our analysis into a methodology for automatically assessing the presence of security issues in workflows.

We implemented the methodology in GHAST (GitHub Actions Security Tester).
GHAST runs on a project source code, taking advantage of the \textit{Sunset} SSC security framework to retrieve the SSC.
Then, it extracts and analyzes the GitHub Actions workflows applying the security checks designed in the methodology.

Using GHAST against 50 open source projects produced relevant experimental results regarding security issues.
We analyzed the results, providing an overview of the current security landscape of GHA workflows.
We identified and manually revisioned the security vulnerabilities discovered through \toolname{}. 

For future work, we plan to deepen the analysis of third-party workflow to \textit{i)} better understand their involvement in the SSC, \textit{ii)} extend GHAST with automatic security assessment of third-party workflow code.

\printbibliography
\end{document}